\documentclass{natureTEST}

\usepackage{graphicx}
\usepackage{bm}
\usepackage{amsmath,amssymb}
\usepackage{xspace}
\usepackage[rightcaption]{sidecap}

\usepackage{textcomp}

\usepackage{lineno} 
\usepackage[hidelinks]{hyperref}

\usepackage[percent]{overpic} 
\usepackage{xcolor} 
\usepackage{upgreek}
\usepackage{ulem}

\title{The low-energy Goldstone mode in a trapped dipolar supersolid}
\author{Mingyang Guo$^{\ast, 1}$, Fabian B\"{o}ttcher$^{\ast, 1}$, Jens Hertkorn$^{\ast, 1}$, Jan-Niklas Schmidt$^1$, Matthias Wenzel$^1$, Hans Peter B\"{u}chler$^2$,\\ Tim Langen$^1$ \& Tilman Pfau$^1$\\
$^{\ast}$These authors contributed equally to this work.}

\begin{document}

\maketitle

\begin{affiliations}
	\item 5. Physikalisches Institut and Center for Integrated Quantum Science and Technology, Universit\"at Stuttgart, Pfaffenwaldring 57, 70569 Stuttgart, Germany
	\item Institute for Theoretical Physics III and Center for Integrated Quantum Science and Technology, Universit\"at Stuttgart, Pfaffenwaldring 57, 70569 Stuttgart, Germany
\end{affiliations}

\begin{abstract}

A supersolid is a counter-intuitive state of matter that combines the frictionless flow of a superfluid with the crystal-like periodic density modulation of a solid\cite{leggett1970can,boninsegni2012colloquium}. Since the first prediction in the 1950s\cite{gross1957unified}, experimental efforts to realize this state have focussed mainly on Helium, where supersolidity remains elusive\cite{chan2013overview}. Recently, supersolidity has also been studied intensively in ultracold quantum gases, and some of its defining properties have been induced in spin-orbit coupled Bose-Einstein condensates (BECs)\cite{Li2017a} and BECs coupled to two crossed optical cavities\cite{leonard2017monitoring, leonard2017supersolid}. However, the periodicity of the crystals in both systems is fixed to the wavelength of the applied periodic optical potentials. Recently, hallmark properties of a supersolid -- the periodic density modulation and simultaneous global phase coherence -- have been observed in arrays of dipolar quantum droplets\cite{Bottcher2019transient, Tanzi2019observation, Chomaz2019long}, where the crystallization happens in a self-organized manner due to intrinsic interactions. In this letter, we prove the genuine supersolid nature of these droplet arrays by directly observing the low-energy Goldstone mode. The dynamics of this mode is reminiscent of the effect of second sound in other superfluid systems\cite{atkins2014liquid, sidorenkov2013second} and features an out-of-phase oscillation of the crystal array and the superfluid density. This mode exists only due to the phase rigidity of the experimentally realized state, and therefore confirms the genuine superfluidity of the supersolid.

\end{abstract}

\begin{figure}[t]
	\centering
	\begin{overpic}[width=0.45\textwidth]{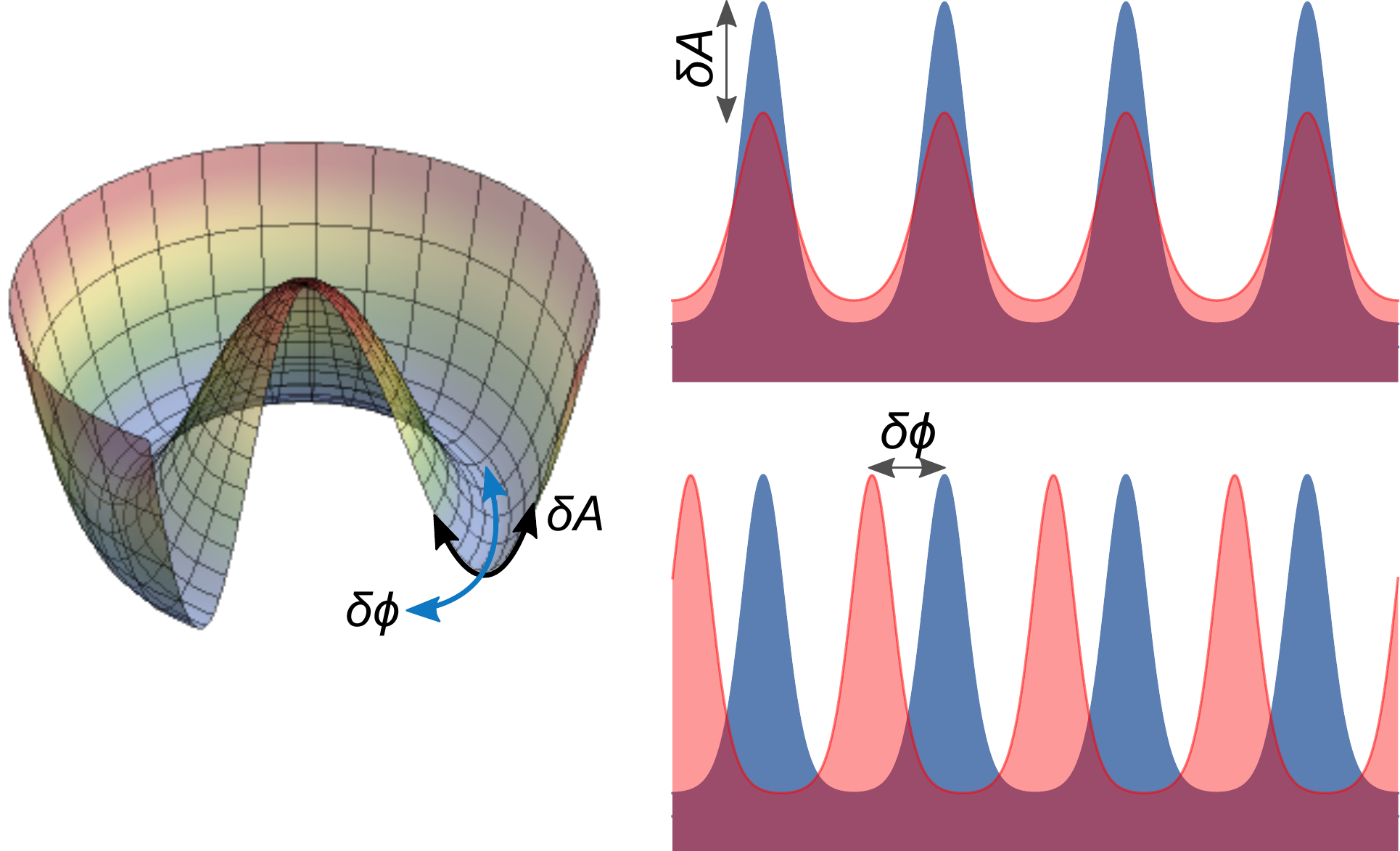}
	\put(3,59){\textbf{a}} \put(45,59){\textbf{b}} \put(45,29){\textbf{c}}
	\end{overpic}
	\caption{\textbf{Goldstone and Higgs modes arising from the broken continuous translational symmetry in an infinite supersolid.} \textbf{a}, Effective potential of an ordered phase as a function of the complex order parameter. Higgs and Goldstone modes correspond to amplitude $\delta A$ and phase $\delta \phi$ modulation of the order parameter $\Psi=Ae^{i\phi}$. The gapless Goldstone mode leads to a degeneracy of the ground state. \textbf{b} \& \textbf{c}, Illustration of the Higgs and Goldstone modes for a broken continuous translational symmetry in an infinite system, for which the modes correspond to an amplitude modulation and a spatial shift of the crystal structure.}
	\label{figure1}
\end{figure}

Symmetry breaking is a crucial concept for describing phase transitions in particle\cite{nambu1961dynamical} and condensed matter physics\cite{griffin1993excitations,anderson1958random}. A spontaneous symmetry breaking occurs when the Hamiltonian of a system is invariant with respect to a certain symmetry while the equilibrium ground state is not. An additional order parameter is therefore necessary to describe the system. A broken continuous symmetry, like e.g. translational invariance or the $U(1)$ symmetry associated with particle number conservation, leads to two types of collective excitations called the Goldstone\cite{goldstone1961field} and Higgs\cite{higgs1964broken} modes. These modes can usually be identified from the resulting effective potential that has the shape of a mexican hat with respect to the order parameter\cite{footnote1}. A schematic example is shown in Fig.~\ref{figure1}\textbf{a}. In this effective potential the gapless Goldstone\cite{goldstone1961field} and the gapped Higgs\cite{higgs1964broken} mode correspond to the phase and amplitude modulation of the complex order parameter at long wavelengths, respectively. In Fig.~\ref{figure1}\textbf{b}~{\&}~\textbf{c}, we schematically show the Goldstone and Higgs mode after breaking the continuous translational symmetry for an infinite system, in which they correspond to a spatial shift (phonons) and an amplitude oscillation of the periodic density modulation. Both Goldstone and Higgs modes have successfully been observed using spectroscopic methods in various platforms including superfluid helium\cite{yarnell1958energy}, solid-state systems\cite{sooryakumar1980raman, littlewood1981gauge, ruegg2008quantum} and ultracold quantum gases\cite{stamper1999excitation, endres2012higgs, hoinka2017goldstone, leonard2017monitoring, behrle2018higgs}.  


\begin{figure*}[!t]
	\centering
	\begin{overpic}[width=0.7\textwidth]{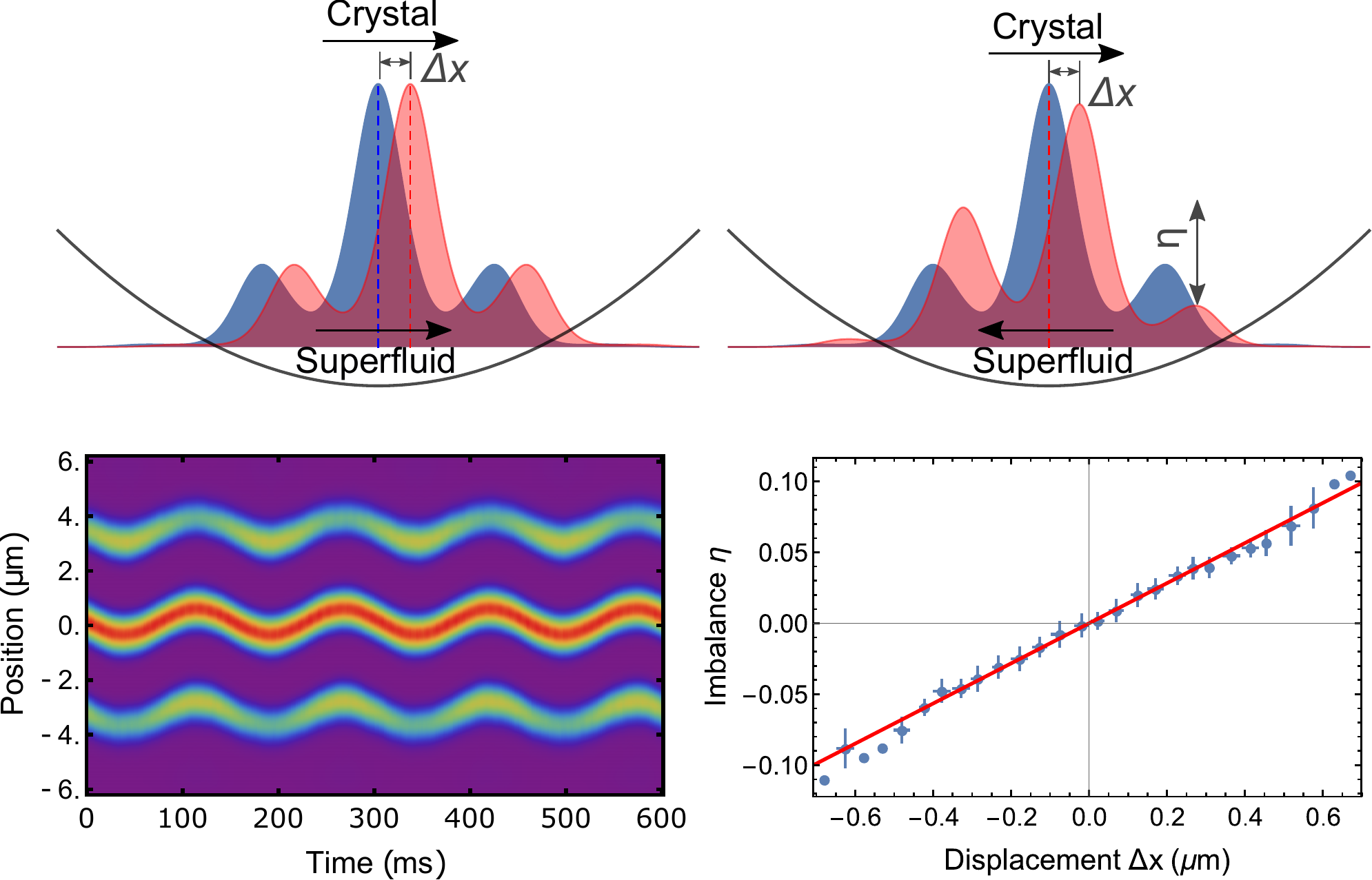}
	\put(3,59){\textbf{a}} \put(55,59){\textbf{b}} \put(3,33){\textbf{c}} \put(55,33){\textbf{d}}
	\end{overpic}
	\caption{\textbf{Goldstone modes in a harmonic potential.} \textbf{a} \& \textbf{b}, The in-phase and out-of-phase Goldstone modes in a harmonic trap in the long-wavelength limit. The arrows indicate the spatial displacement $\Delta x$ of the droplet array and the flow of the superfluid density. Vertical dashed lines indicate the center-of-mass (COM) position of the whole cloud. The in-phase mode corresponds to the COM dipole oscillation with the trap frequency, while the out-of-phase mode maintains the COM due to the counterflow of the droplet array and the superfluid background.  As a result this mode has a very low energy. \textbf{c}, Simulated dynamics of the out-of-phase Goldstone mode for the ground state, an array of three droplets. The spatial motion of the droplet array is compensated by a superfluid flow of atoms in the opposite direction in order to maintain the COM. \textbf{d}, Numerically predicted correlation between the imbalance $\eta$ and the displacement $\Delta x$ of the droplet array, together with a linear fit. The shown error bars indicate the uncertainty of the fit used to extract $\eta$ and $\Delta x$.}
	\label{figure2}
\end{figure*}
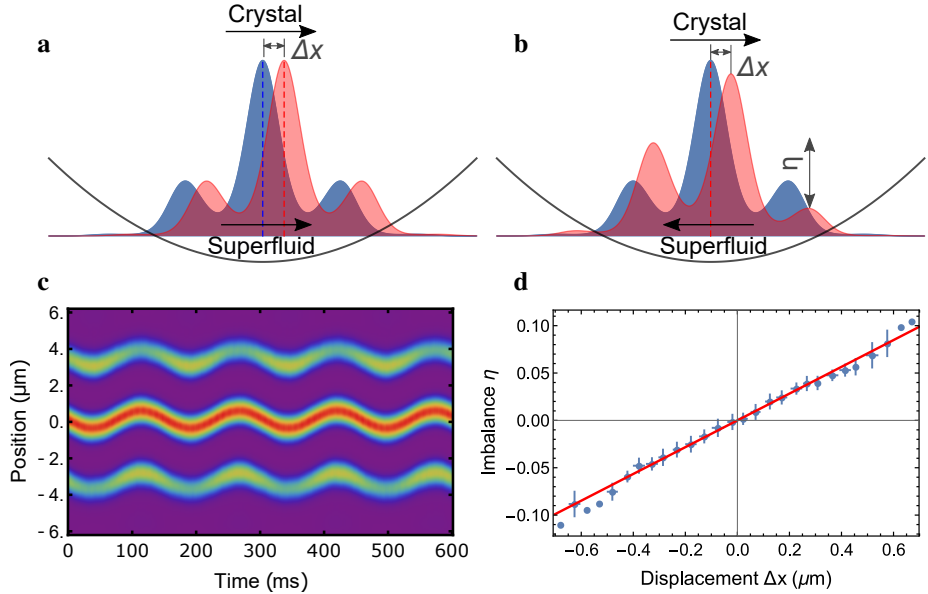

A quantum mechanical ground state that simultaneously breaks the global $U(1)$ symmetry and the continuous translational symmetry is the long-sought supersolid. One promising system to realize supersolidity are dipolar quantum gases, which inherently feature a preferred length scale for a periodic density modulation due to their roton-like dispersion relation\cite{Santos2003,Chomaz2018, petter2019probing}. For increasingly dominant dipolar interactions, the roton mode softens and finally becomes imaginary, at which point mean-field theory predicts a collapse. 
However, this collapse is prevented by beyond mean-field effects leading to stable quantum droplets\cite{Kadau2016, Ferrier-Barbut2016, Ferrier-Barbut2016a, Wenzel2017, Ferrier-Barbut2018}. Supersolid ground states formed by arrays of these droplets have then been proposed for suitably chosen combinations of confinement and scattering length\cite{roccuzzo2019supersolid, Wenzel2017}. Following this, arrays of quantum droplets featuring supersolid properties -- specifically the crystalline density modulation and global phase coherence -- have recently been realized experimentally\cite{Bottcher2019transient, Tanzi2019observation, Chomaz2019long}. Distinct from spin-orbit coupled BECs\cite{Li2017a} and BECs coupled to two crossed optical cavities\cite{leonard2017monitoring, leonard2017supersolid}, spatial ordering of these droplet arrays arises from intrinsic interactions between the particles, and therefore the crystal can support phonon modes. However, a definite proof of phase rigidity, and therefore genuine superfluidity, is still missing and required in order to verify the supersolid nature of the system.

In this letter, we describe the experimental realization of a state, that simultaneously shows all three necessary hallmarks of a supersolid -- a periodic density modulation, global phase coherence and phase rigidity. In our work, the phase rigidity is proven by studying the low-energy Goldstone mode of the system that arises due to the two broken symmetries. The observed low-energy Goldstone mode features an out-of-phase oscillation between the droplet array and the superfluid density, involving Josephson-like dynamics between the droplets and therefore highlighting the phase rigidity of the state. 

For an infinite supersolid droplet array that simultaneously breaks the global $U(1)$ symmetry and the continuous translational symmetry, we can define the corresponding two complex order parameters as the condensate wavefunction $\Psi(k=0)$ and the periodic density modulation $\varrho(k = \frac{2\pi}{d}) = \mathcal{F}\left[\left| \Psi(x) \right|^2\right]$ with a wavelength $d$. Accompanying the two order parameters, there are two branches of Goldstone and Higgs modes. This leads to a complex excitation spectrum\cite{Saccani2012, Macri2013}, featuring two kinds of sound, one corresponding to the compressibility of the supersolid and one corresponding to the superfluid stiffness. Correspondingly for a trapped supersolid droplet array with a finite size, the two Goldstone modes consist of an in-phase combination of phonons in the BEC and the crystal part of the supersolid (Fig.~\ref{figure2}\textbf{a}), and an out-of-phase oscillation of the two (Fig.~\ref{figure2}\textbf{b}). The in-phase mode is exactly the center-of-mass (COM) dipole oscillation of the whole cloud at the trap frequency. However, the out-of-phase mode maintains the COM by a precise interplay of the crystal motion and the superfluid counterflow, which can have a significantly lower energy. In contrast to the low-energy Goldstone mode, the Higgs mode in our system is expected to be strongly damped and therefore only plays a negligible role in our observations.

To thoroughly explore the two modes in trapped droplet arrays, we implement Bogliubov-de Gennes theory based on the extended Gross-Pitaevskii equation (eGPE) including the relevant beyond mean-field corrections\cite{ronen2006bogoliubov, wilson2010critical, roccuzzo2019supersolid}. In the supersolid region, both the in- and out-of-phase mode are found at the long-wavelength limit that correspond to the sample size (see Methods). As expected, the energy of the in-phase mode exactly corresponds to the trap frequency, while the out-of-phase Goldstone mode has an excitation energy much lower than the trap frequency. The simulated time evolution of the out-of-phase mode in a supersolid array is shown in Fig.~\ref{figure2}\textbf{c}. Starting from the symmetric three-droplet ground state, the droplet array moves to one side. This movement of the droplet array is accompanied by a superfluid atom transport that redistributes the atoms in order to maintain the COM. For large displacements, which are realized for larger excitation amplitudes, we observe an oscillation between a three- and four-droplet state (see Methods). For each time step we confirm that the COM stays unchanged within our simulation resolution. Furthermore, from this simulation we can obtain the oscillation frequency of 5.6~Hz for this low-energy mode for the parameters used in the simulation of Fig.~\ref{figure2}\textbf{c}, in agreement with the Bogoliubov result. 

In order to maintain the COM, the superfluid flow characterized by the droplet imbalance $\eta$ of the droplet array and the crystal displacement $\Delta x$ must satisfy a certain relation. In an array of three droplets, we define the imbalance as $\eta=(N_1-N_3) / (N_1+N_2+N_3)$ with $N_i$ the atom number in the $i$th droplet, numbered from left to right, and the displacement $\Delta x$ as the arithmetic mean of the positions of the three droplets relative to the COM of the whole cloud, also including a thermal background. Fig.~\ref{figure2}\textbf{d} shows the strong correlation between $\eta$ and $\Delta x$ extracted from the numerical simulation shown in Fig.~\ref{figure2}\textbf{c}, which is found to be the same for different excitation amplitudes. More importantly, the correlation is also robust against a small variation of the scattering length, corresponding to different fractions of the superfluid background. 
Therefore, the existence of such a correlation acts as a clear signature for the out-of-phase Goldstone mode, and thus proves supersolidity of the system.


In order to experimentally observe this low-energy Goldstone mode, we prepare ultracold quantum gases of $^{162}$Dy in an elongated optical dipole trap with trap frequencies $\omega=2\pi~[30(1), 89(2), 108(2)]~\text{Hz}$ and the magnetic field along the \^y axis (see Methods). Depending on the final scattering length, the prepared cloud can be in the normal BEC phase ($>100~a_0$), an incoherent droplet array ($<96~a_0$), or a coherent droplet array in between.
After reaching the final scattering length, we let the cloud evolve and equilibrate for 15~ms before we probe it using an in-situ phase-contrast imaging. 

As the oscillation period of the out-of-phase mode is much longer than the droplet lifetime, which is limited by three-body losses\cite{Bottcher2019transient}, it is currently not feasible to use a standard spectroscopic method\cite{stamper1999excitation, petter2019probing} to detect the Goldstone mode. However, the low excitation energy of the out-of-phase Goldstone mode leads to its excitation within the energy bandwidth of the dynamical formation process. As a result, the mode should always be excited in the prepared samples  with a different phase due to experimental imperfections, such as the non-adiabatic ramping of the scattering length, technical noise, and thermal fluctuations. We therefore repeat the experiment many times at each scattering length, in order to statistically map out the correlation between the imbalance $\eta$ and the crystal displacement $\Delta x$. 

\begin{figure}[!t]
	\centering
	\begin{overpic}[width=0.4\textwidth]{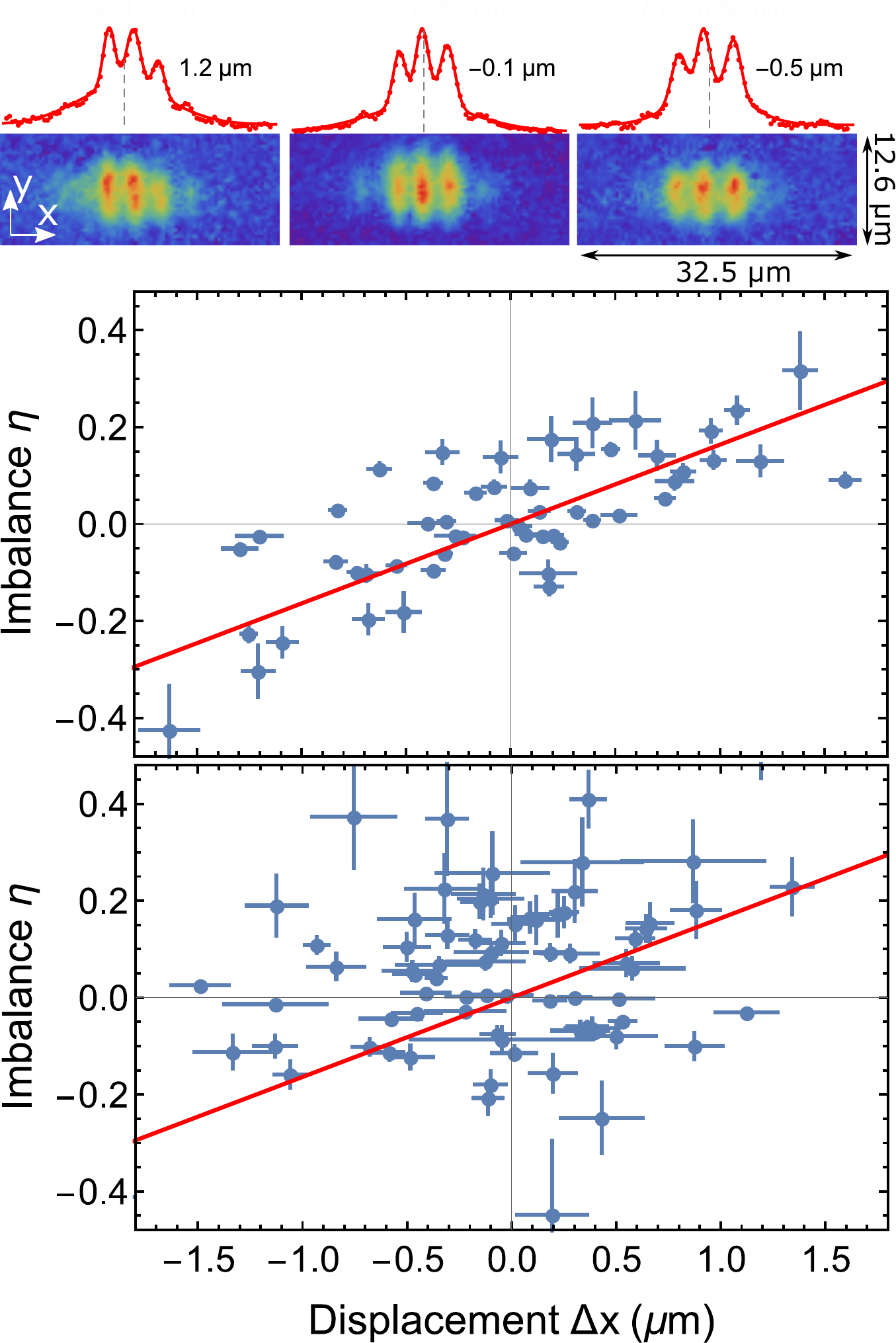}
	\put(0,97){\textbf{a}} \put(0,78.5){\textbf{b}} \put(0,43){\textbf{c}}
	\end{overpic}
	\caption{\textbf{Experimental correlation between $\mathbf{\eta}$ and $\mathbf{\Delta x}$.} \textbf{a}, Exemplary in-situ images with different displacements of the droplet array relative to the total COM of the whole cloud (dashed line). When the droplet array shifts to the right (left) with a positive (negative) $\Delta x$, atoms flow to the left (right) resulting in higher atom number on the left (right) side, in order to keep the COM unchanged. The integrated densities together with the fits (red curves) are also shown.
	\textbf{b} \& \textbf{c}, Experimental correlations in the supersolid ($97.6~a_0$) and isolated droplet ($91.2~a_0$) region. The error bars correspond to the uncertainty of the fits used to extract the position and atom number of each individual droplet. The red curve is the theoretical prediction for a supersolid state without any free parameters as shown in Fig.\ref{figure2}\textbf{d}. A correlation is clearly demonstrated in the supersolid region, while such a correlation is missing in the isolated droplet phase. 
}
\label{figure3}
\end{figure}

After post selection on the total atom number (see Methods), $\Delta x$ and $\eta$ are extracted for each cloud by fitting the in-situ density distributions with a sum of four Gaussian functions, corresponding to a BEC or thermal background and three droplets. Some example images with different imbalances and displacements $\Delta x$ for a scattering length of 97.6~$a_0$ are shown in Fig.~\ref{figure3}\textbf{a} to visualize how the imbalance changes with respect to the spatial shift of the droplets. Fig.~\ref{figure3}\textbf{b} shows the measured correlation between $\eta$ and $\Delta x$ for the observed three-droplet states in the supersolid region, where a strong correlation is observed. More importantly, the correlation perfectly coincides with the theoretical prediction (red curve in Fig.~\ref{figure3}\textbf{b}) without any free parameters. We interpret the spread in the data around the theoretical correlations as imperfections in the extraction of the imbalance and the displacement, as well as small excitations of higher lying modes. The existence of the correlation proves the presence of the low-energy Goldstone mode, and therefore supersolidity of the system. For comparison, one example of isolated droplet arrays at 91.2~$a_0$ is shown in Fig.~\ref{figure3}\textbf{c}, where the correlation is missing, resulting from the lack of superfluid counterflow due to the vanishing superfluid fraction.

\begin{figure}[!t]
	\centering
	\begin{overpic}[width=0.4\textwidth]{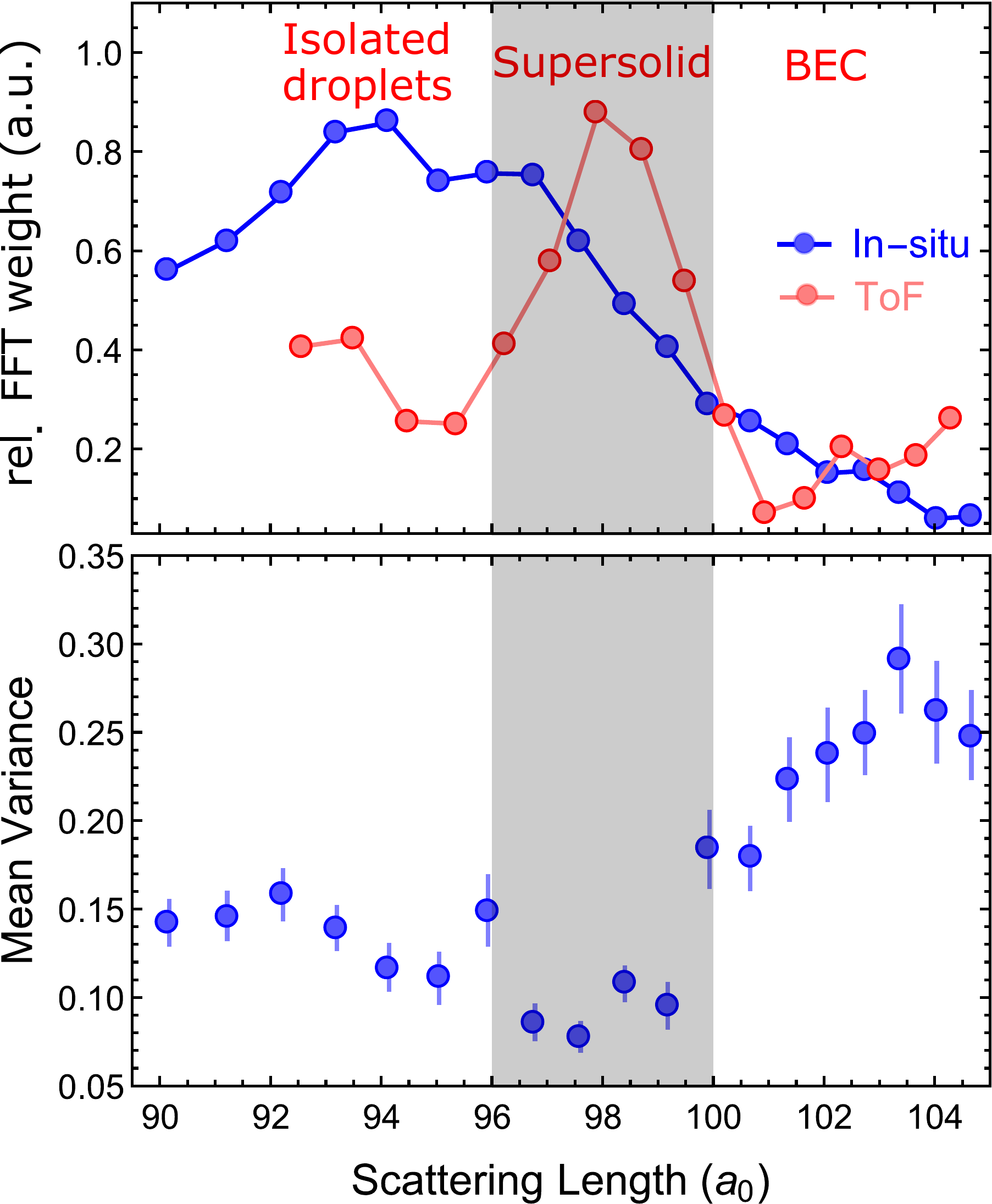}
	\put(0,98){\textbf{a}} \put(0,52){\textbf{b}}
	\end{overpic}
	\caption{\textbf{The ($\eta - \Delta x$) correlation across the phase diagram.} \textbf{a}, Experimental signature of the phase diagram for our trap geometry. We can identify the different regions by evaluating the in-situ density modulation (blue, characterized by the corresponding Fourier weight at the modulation wavelength) and the global phase coherence (red, characterized by the Fourier weight of in time-of-flight interference patterns)\cite{Bottcher2019transient}. The shadowed area is the determined coherent droplet region. \textbf{b}, Mean variance of the experimental data with respect to the theoretical correlation curve for the supersolid state. Coinciding exactly with the phase-coherent region in \textbf{a} is the range where we observe the smallest variance, and therefore the strongest correlation, which indicates the existence of the low-energy Goldstone mode. The errors bars indicate the standard error of the experimental data with respect to the obtained correlation from the theoretical simulations.
}
\label{figure4}
\end{figure}

To quantify the range of the contact interaction for which we can observe supersolid arrays of quantum droplets, we calculate the variance of the experimental data with respect to the obtained correlation from the theoretical simulations. A small variance therefore is evidence for the existence of the low-energy Goldstone mode. The measured variance of the data compared to the theory across the explored region is shown in Fig.~\ref{figure4}\textbf{b}. We can clearly see a region with smaller variance, and therefore stronger correlation, which coincides exactly with the region of global phase coherence (shadowed area in Fig.~\ref{figure4}\textbf{a}). The latter can independently be determined from time-of-flight interference\cite{Bottcher2019transient}. In the BEC phase, some clouds are also detected to be density modulated, which may result from excitations of the roton mode, which close to the phase transition also has a small excitation energy due to its softening. However, the correlation is also missing in this region. As a theory independent test, we analyze the $\eta - \Delta x$ correlation data with a linear fit and confirm that only in the supersolid region the displacement can be compensated by a superfluid flow, and therefore an imbalance of the droplet array (see Methods).

In conclusion, we have studied the low-energy Goldstone mode in a trapped supersolid droplet array of dipolar dysprosium atoms which enables the simultanous detection of phase coherence and phase rigidity. The out-of-phase mode features a counterflow of the crystal-like droplet array and the superfluid density, leading to a robust correlation between the imbalance and the displacement of the droplets. This mode therefore directly connects to the broken $U(1)$ symmetry and the continuous translational symmetry, highlighting the supersolid nature of the coherent arrays of dipolar quantum droplets. An extension to our work would be the observation of other collective excitations, especially the Higgs modes, which require additional symmetries to suppress their decay into other lower-energy excitations\cite{littlewood1981gauge}. Another promising direction is realizing a supersolid state with larger droplet numbers, or even two-dimensional supersolid arrays\cite{baillie2018droplet}, where an additional spatial symmetry is broken, leading to an even more complex excitation spectrum.


Note: During the preparation of this manuscript, we became aware of a related complimentary study of higher lying collective modes, showing a bifurcation upon the transition from a regular BEC to the supersolid droplet array \cite{newModugno}.

\begin{methods}
\setcounter{figure}{0}
\renewcommand{\figurename}{Supplementary Figure}
\renewcommand{\thefigure}{S\arabic{figure}}

\paragraph{Sample preparation and experimental details.} The complete experimental sequence is described in detail in our previous publications\cite{Kadau2016,Bottcher2019transient}. In short, we prepare a quasi-pure BEC of $^{162}$Dy in a crossed optical dipole trap formed by two laser beams at 1064 nm. The degenerate cloud typically contains of $4\times10^4$ atoms at a temperature below 10~nK. After evaporation, the trap is reshaped within 20\,ms to the final trap geometry with trap frequencies of $\omega~=~2\pi~[30(1),~89(2),~108(2)]~\text{Hz}$. For the presented measurements the magnetic field is orientated along the \^y-direction. Subsequently, we tune the contact interaction from the background scattering length $a_{bg}=140(20)~a_0$\cite{Tang2015a,Tang2016,Tang2018} to approximately 112~$a_0$ by ramping the magnetic field closer to the double Feshbach resonances near 5.1~G\cite{Baumann2014,Bottcher2019transient,Bottcher2019quantum}. To reach the droplet region, the magnetic field is further linearly ramped to the final scattering length in the range of 90~$a_0$ and 105~$a_0$ in 30~ms. We then let the samples evolve for 15~ms in order to allow the quantum droplet arrays to form and equilibrate. Subsequently we probe the atomic clouds with our in-situ phase-contrast imaging, which is performed along the \^z axis using a microscope objective with a numerical aperture of 0.3. Our resolution of about 1~$\upmu$m allows us to distinguish nearby droplets that are separated by $\sim$3~$\upmu$m. By fitting each individual droplet with a Gaussian, we can extract the atom number in each individual droplet, as well as the position of the droplet to a higher precision than our imaging resolution. 

To verify the range for which we observe phase-coherent droplet arrays, we implement a time-of-flight interference sequence similar to our previous work\cite{Bottcher2019transient}. For this we ramp up the scattering length within 100\,$\upmu$s to $\sim a_{bg}$, in order to accelerate the expansion and then release the atoms from the trap. Due to the changed geometry the expansion time is now limited to 7.2\,ms compared to our previous publication. Nonetheless, we can observe clear interference patterns and are able to distinguish between phase-coherent and incoherent droplet regimes. 

Notably, for samples with smaller scattering lengths, which means for magnetic fields closer to the Feshbach resonances, the clouds suffer from more severe three-body loss. The black points in Fig.~\ref{sup1} indicates the corresponding average atom number in the experiment for each scattering length. For our evaluation we post-select realizations in a range of $\pm 15\%$ around the average atom number at each scattering length are selected in the statistical evaluation from which the correlation is extracted. The experimental realizations mostly consist of three or four droplets. We analyze these two cases separately after distinguishing them with multiple Gaussian fits. Although the COM is experimentally determined by the whole image, it is dominated by the condensate and thermal background within the BEC and coherent droplet region. For isolated droplets we also observe a thermal background, which then acts as the main contribution to the determined COM of the whole cloud, that we use as a reference for the calculation of the displacement. 

For the range of scattering lengths studied in this work, our experimentally determined stability of the magnetic field leads to an uncertainty of $\sim$1~$a_0$, while the calibration of the positions and widths of the double Feshbach resonances, that we use to tune the scattering length, results in an uncertainty of $\sim$4~$a_0$. On top of this uncertainty, there is an overall systematic uncertainty due to the absolute value of the background scattering length\cite{Tang2015a,Tang2016,Tang2018}, which has so far not been measured to high precision. This leads to an uncertainty of all calculated scattering lengths that is on the order of 15\%.

\begin{figure}[t]
	\centering
	\begin{overpic}[width=0.45\textwidth]{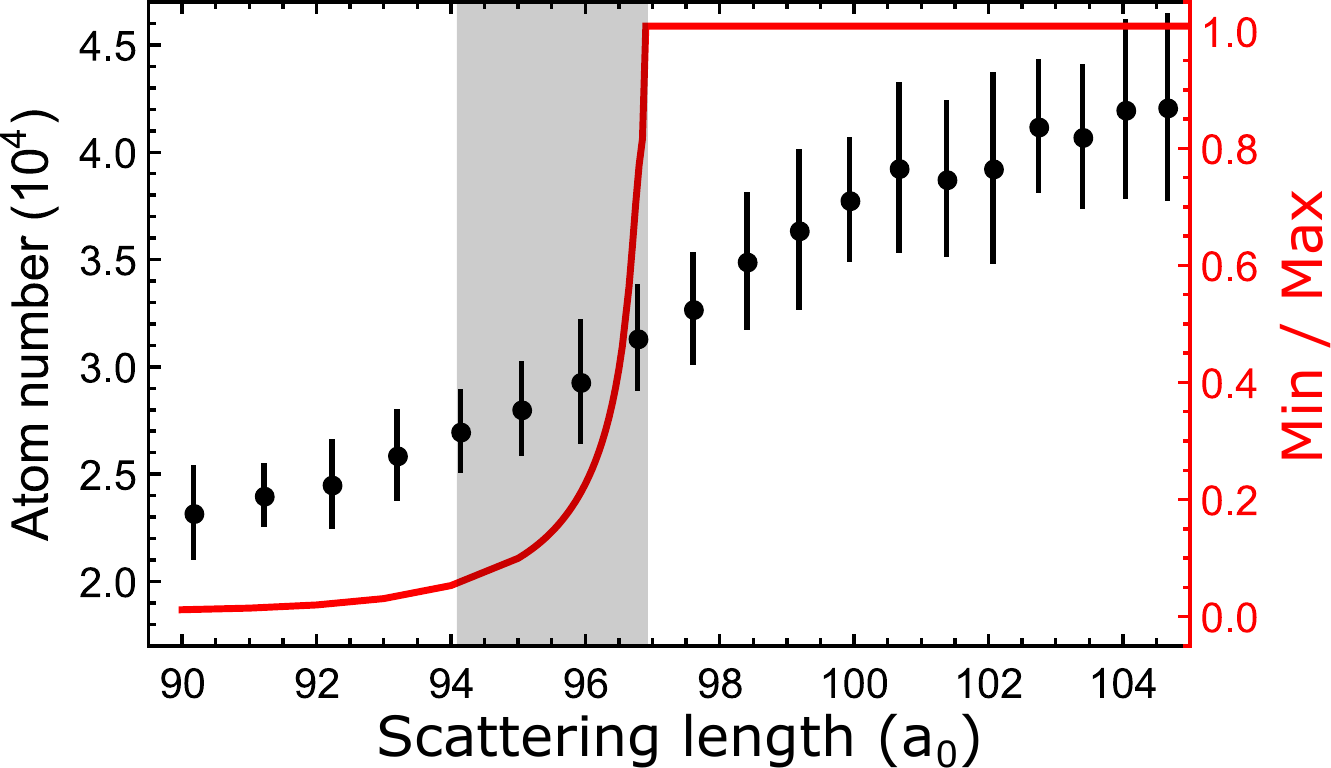}
	\end{overpic}
	\caption{\textbf{Theoretical phase boundaries.} As an indicator of the phase coherence, we show the ratio of the first minimum in the density compared to the central droplet peak density as an indicator of the overlap between the droplets\cite{Bottcher2019quantum} of the calculated density profile of the ground state in red. As in the experiment, three regions are identified and the coherent region locates between 94~$a_0$ and 97~$a_0$, shifting $\sim$3~$a_0$ from the experimentally obtained phase diagram. These simulations are done for an atom number of $30\times10^3$. The black points indicate the measured average atom number in the experiment at each scattering length. 
	}
	\label{sup1}
\end{figure}

\paragraph{Theory: eGPE simulations and Bogoliubov excitation spectrum.} Our theory is based on numerical solving the extended Gross-Pitaevskii equation (eGPE), which includes quantum fluctuations as beyond mean-field corrections. We obtain the ground states by performing imaginary time evolution\cite{Wenzel2017} of the eGPE including a harmonic potential with $\omega=2\pi~[30, 90, 110]~\text{Hz}$, similar to the experiment. In this trap we obtain a three-droplet ground state for scattering lengths $a_{\text{s}} \geq 91\,a_0$, as it is shown in Fig.~\ref{sup3}. To identify the phase-coherent droplet arrays, we use the same indicator of the nearest minimum divided by the central maximum as an estimation of the overlap between the droplets like in our previous publication\cite{Bottcher2019transient}. The result is shown in Fig.~\ref{sup1}, where we can succesfully identify the three different regions, with the coherent region being located between $94-97~a_0$. Compared to the experimental data, the obtained coherent region from the simulations is shifted by $3\,a_{0}$, in line with recent results in Erbium showing a similar deviation\cite{petter2019probing, Chomaz2019long}.

\begin{figure}[t]
	\centering
	\begin{overpic}[width=0.48\textwidth]{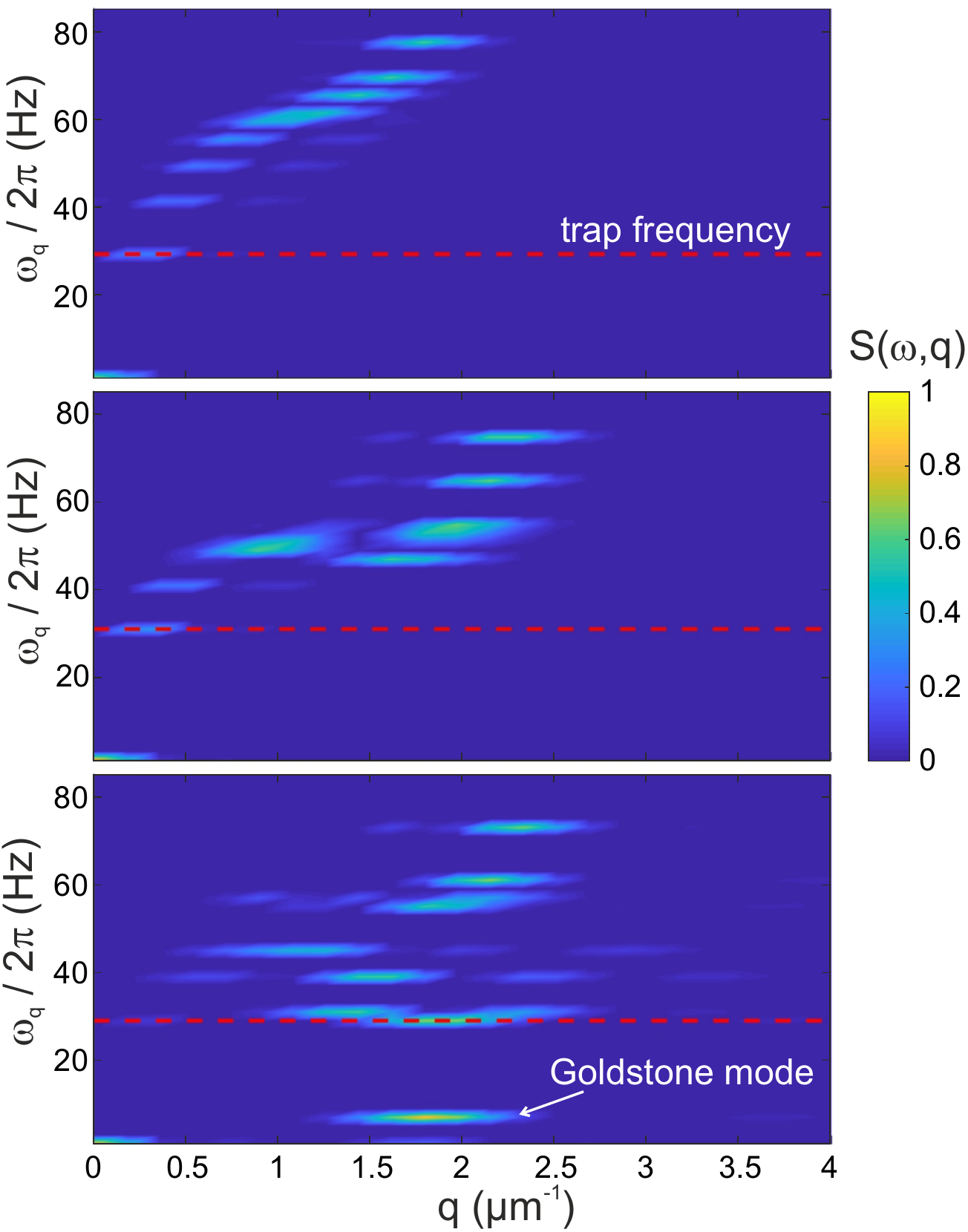}
	\put(63, 96){\textbf{\textcolor{white}{a}}} \put(63, 64){\textbf{\textcolor{white}{b}}} \put(63, 33){\textbf{\textcolor{white}{c}}}
	\end{overpic}
	\caption{\textbf{Dynamic structure factor of the collective excitations.} Calculated structure factor $S(\omega,q)$ in the BEC phase for $a_{\text{s}} = 100\,a_0$ (\textbf{a}) and $a_{\text{s}} = 98\,a_0$ (\textbf{b}) and in the supersolid droplet arrays (\textbf{c}) for $a_{\text{s}} = 96\,a_0$. Due to the finite size of the system, long wavelength modes become discrete and the lowest possible excitation energy is set by the trap frequency (dashed red line). For decreasing contact interaction strengths we can observe the roton minimum emerging, until finally its gap closes and the supersolid appears. In the supersolid regime we can clearly see the low-energy out-of-phase Goldstone mode and the large gap to all the other modes above the trap frequency of 30\,Hz. The color code is normalized to the mode with the highest response across the shown scattering lengths.
	}
	\label{sup2}
\end{figure}

To calculate the excitation spectrum, we follow the Bogoliubov-de Gennes theory by linearizing the eGPE around the ground states \cite{ronen2006bogoliubov}. The resulting Bogoliubov–de Gennes equations are then solved numerically to obtain the excitation modes\cite{ronen2006bogoliubov}. The obtained dispersion relations are exemplarily shown in Fig.~\ref{sup2} for three different scattering lengths. Due to finite sample size, the excitation spectra become discrete with each mode corresponding to a spread out momentum. In order to visualize the spectra we calculate the zero-temperature structure factor $S(\omega,q)$\cite{Brunello2001, Blakie2002} that indicates the dynamic response of the cloud. The higher the amplitude of this structure factor, the stronger is the density response of the system to the corresponding mode. In this work, we ignore the modes with excitation energies higher than $\sim$90~Hz, where excitations along the other trap axes start to play a role and make the spectrum much more complex. 

In the BEC phase (Fig.~\ref{sup2}a), there is only one excitation branch, featuring an increasing excitation energy with increasing momentum $q$. By decreasing the contact interaction strength we can observe the appearance of the the roton mode in Fig.~\ref{sup2}b. Lowering the scattering length further, the roton mode softens and we enter the supersolid regime, where an additional excitation branch with lower energy appears due to breaking of the continuous translational symmetry. This low-energy mode corresponds to the out-of-phase Goldstone mode, whose correlation we directly observe in the experiment. Close to the phase transition from BEC to supersolid, the calculations clearly reveal a large energy splitting between this low energy mode and any other collective mode, especially the in-phase COM mode at the trap frequency. For smaller contact interaction strength we reach the regime of isolated droplets, where the low-energy out-of-phase decreases in energy. At the same time, we observe the emergence of a clear periodicity in the excitation spectrum due to the underlying crystal structure. 


\paragraph{Dynamics of the out-of-phase Goldstone mode.} From the numerically solved Bogoliubov-de Gennes equations, we can also obtain the phase pattern of each excitation mode. In Fig.~\ref{sup3} we show the calculated  $1D$-line cut through the $3D$ phase pattern of the out-of-phase Goldstone mode, together with a density cut through the calculated density profile of the three-droplet ground state. Similar as in Josephson dynamics, the phase gradient is directly proportional to the particle flow. Distinct from the phase pattern of the in-phase dipole mode which is a constant phase gradient over the cloud, the out-of-phase mode has a step-wise phase pattern, with each step coinciding with a droplet. While the whole BEC background thus experiences a phase gradient with a particular direction, the droplets always experience a gradient in the opposite direction. This leads to the counterflow between the BEC background and the crystal that characterizes this mode. By numerically imprinting the phase pattern of a specific collective excitation onto the ground state, we can directly simulate the excitation dynamics of each individual mode by performing a real-time evolution of the eGPE. Doing this for the out-of-phase Goldstone mode, we obtain the dynamical time evolution of this low-energy mode, as it is shown in Fig.~\ref{figure2}\textbf{c} of the main text.

\begin{figure}[!t]
	\centering
	\includegraphics[width=0.45\textwidth]{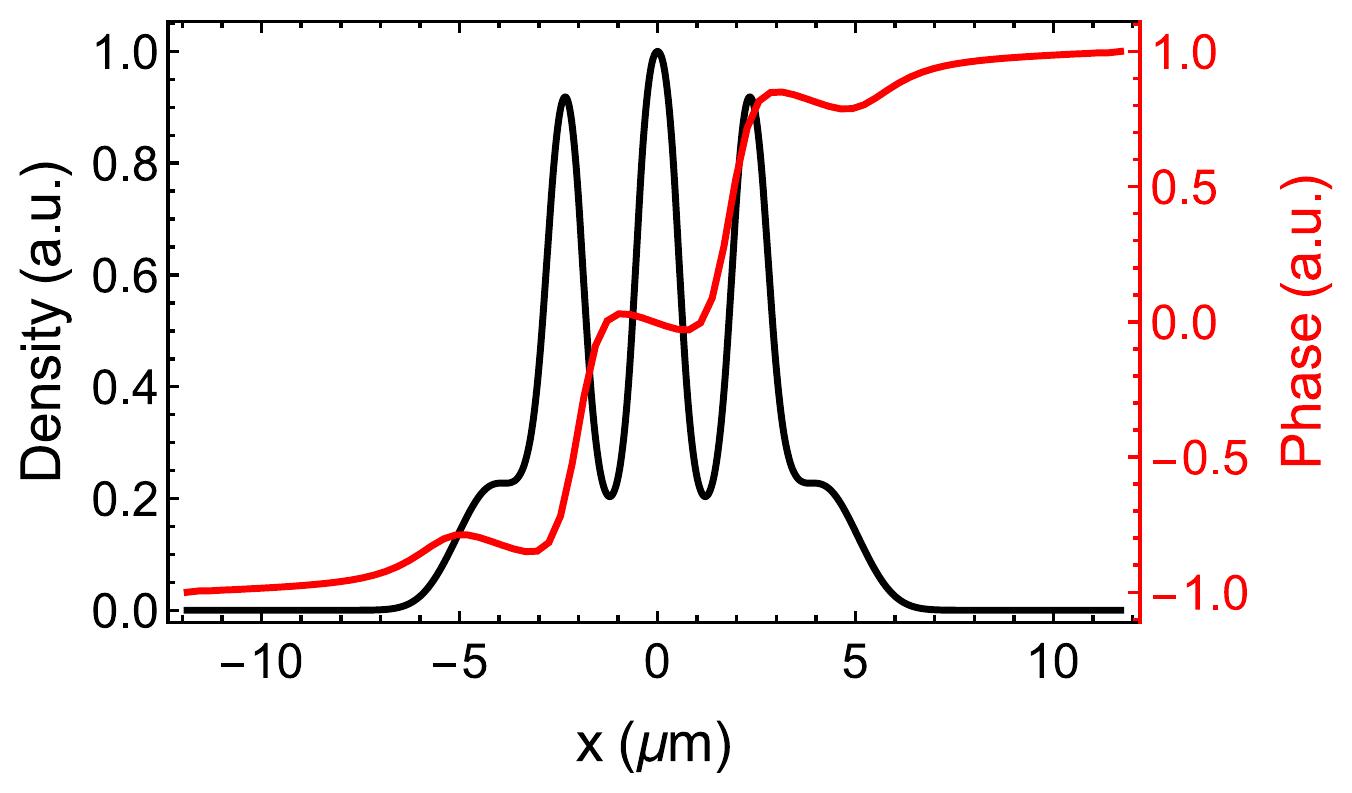}
	\caption{\textbf{Phase pattern of the Goldstone mode.} Density cut through the three-droplet ground state (black) along the \^x axis and the phase pattern (red) corresponding to the low-energy Goldstone mode. 
	}
	\label{sup3}
\end{figure}


We simulate the real-time dynamics of the system for different scattering lengths in the supersolid range. Decreasing the scattering length we observe an increase of the oscillation period, in line with the results obtained from the calculated dispersion relations. This decrease of the oscillation frequency is accompanied by a decrease in the oscillation amplitude. Looking at the correlation between the imbalance $\eta$ and displacement $\Delta x$ that we use as an indication for the presence of the low-energy Goldstone mode, we observe that it remains unchanged within the uncertainty of our evaluation, even though the oscillation frequency changes by nearly doubles in the studied range of the scattering length. 

Using a similar procedure, we have also checked that the correlation is independent of the excitation amplitude. This is the case as long as the oscillation amplitude remains sufficiently small. In our simulations for larger oscillation amplitudes, we can observe that the three-droplet state changes to a four droplet state for large displacements. As an example, we show two simulated real-time evolutions with higher amplitudes of the low-energy mode in Fig.~\ref{SupTimeEvoLareAmp}. The change from a three- to a four-droplet state can happen, because in our particular trap and at the studied scattering lengths, the energy difference between the two states is small. However, the periodic coherent emergence and disappearance of a new droplet on the edges of the system is another clear evidence for the presence of superfluid flow. For very large amplitudes, as it is shown in Fig.~\ref{SupTimeEvoLareAmp}\textbf{b}, the excitation is no longer oscillatory, but traveling in one direction, reminiscent of the Goldstone mode in an infinite system (see Fig.~\ref{figure1}\textbf{c} of the main text).

\begin{figure}[t]
	\centering
	\begin{overpic}[width=0.42\textwidth]{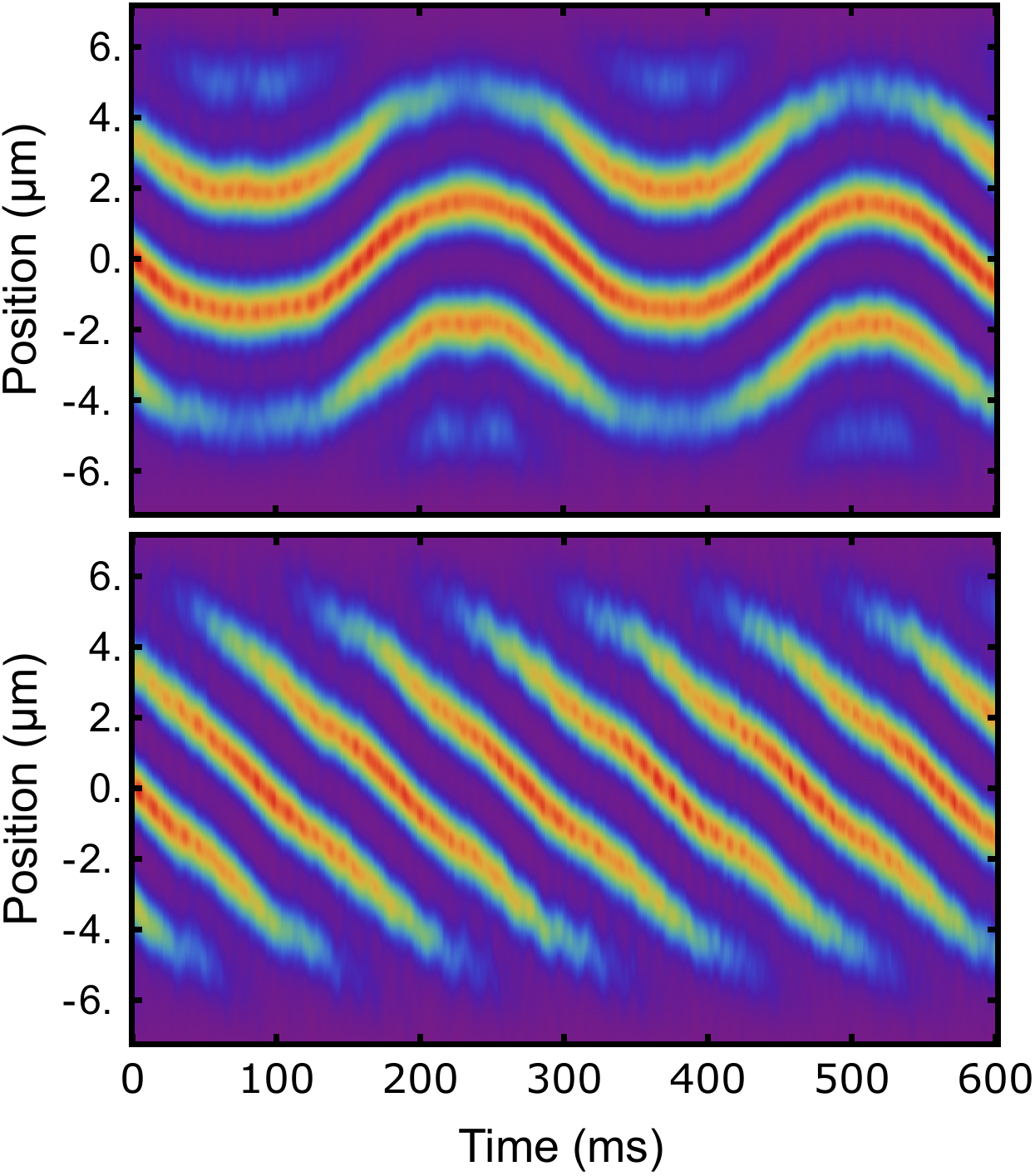}
	\put(1, 97){\textbf{a}} \put(1, 52){\textbf{b}}
	\end{overpic}
	\caption{\textbf{Large amplitude dynamics of the out-of-phase Goldstone mode.} Starting from an array of three droplets, the state can change to a four-droplet state for large excitation amplitudes. From there it either oscillates back and forth between the two (\textbf{a}) or the excitation amplitude is so large that we find that the motion is not even oscillatory anymore, but that the excitation rather travels only in one direction (\textbf{b}).
	}
	\label{SupTimeEvoLareAmp}
\end{figure} 

\paragraph{Evaluation of four-droplet states.}
For large displacements in our dynamical simulations of the Goldstone mode, that happen at a large amplitude of the excitation, the three-droplet ground state can change to a four-droplet state. Since the two states with different droplet number are smoothly connected, we also observe four-droplet states in the experiment, due to strong excitations of the Goldstone mode.

\begin{figure}[t]
	\centering
	\includegraphics[width=0.44\textwidth]{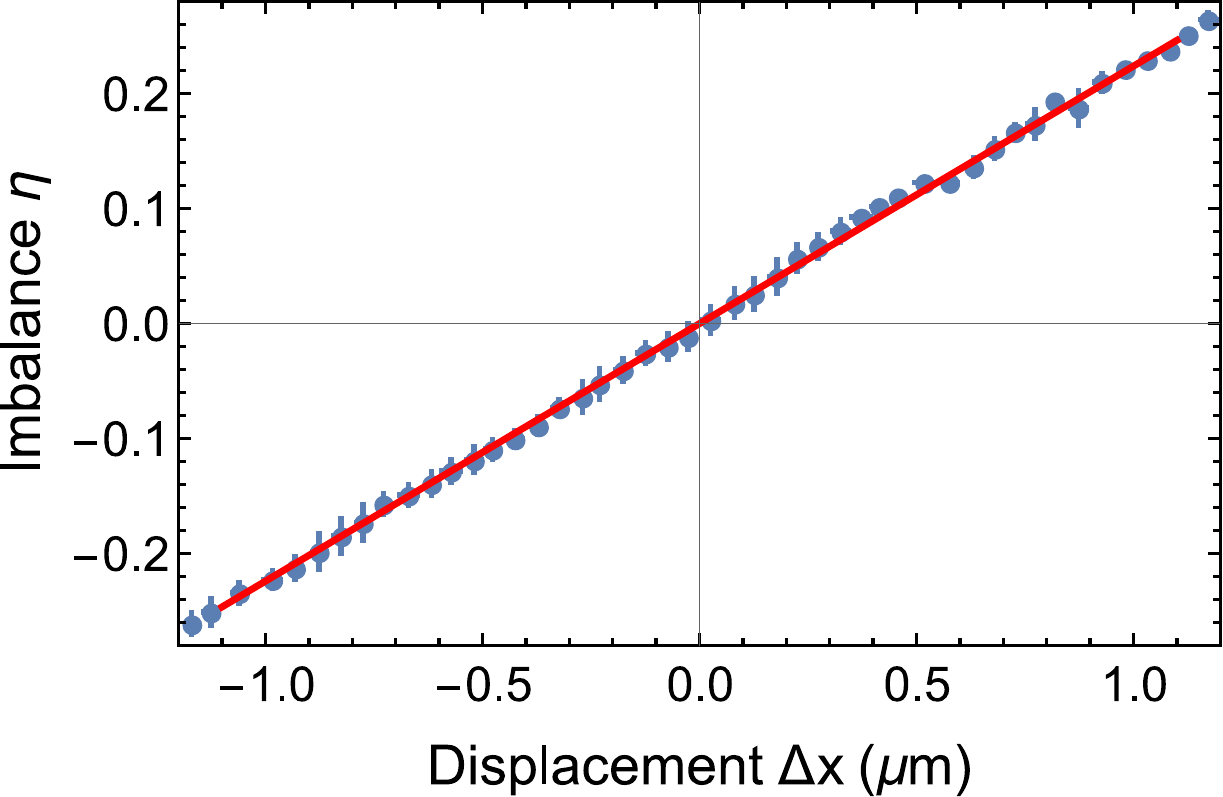}
	\caption{\textbf{Simulated ($\eta - \Delta x$) correlation of the four-droplet dynamic states.} Numerically predicted correlation between the imbalance $\eta$ and droplets displacement $\Delta x$ for the four-droplet states appearing at large excitation amplitudes of the low-energy Goldstone mode (blue points). The red line corresponds to a linear fit.
	}
	\label{sup4}
\end{figure} 

In order to maintain the COM, this mode should again satisfy a certain relation between the imbalance, which we now define as $\eta~=~(N_1~+~N_2~-~N_3~-~N_4)~/~(N_1~+~N_2~+~N_3~+~N_4~)$ for a four-droplet state, and the displacement $\Delta x$, which we again define as the arithmetic mean of all four positions of the individual droplets relative to the COM of the whole cloud. The calculated correlation of a four-droplet state is shown in Fig.~\ref{sup4}. The data in this plot corresponds to the large displacement data that was shown in Fig.~\ref{SupTimeEvoLareAmp}. Again, we obtain a linear dependence, that is robust against variations of the scattering length, as well as the excitation amplitude.

\begin{figure}[t]
	\centering
	\begin{overpic}[width=0.48\textwidth]{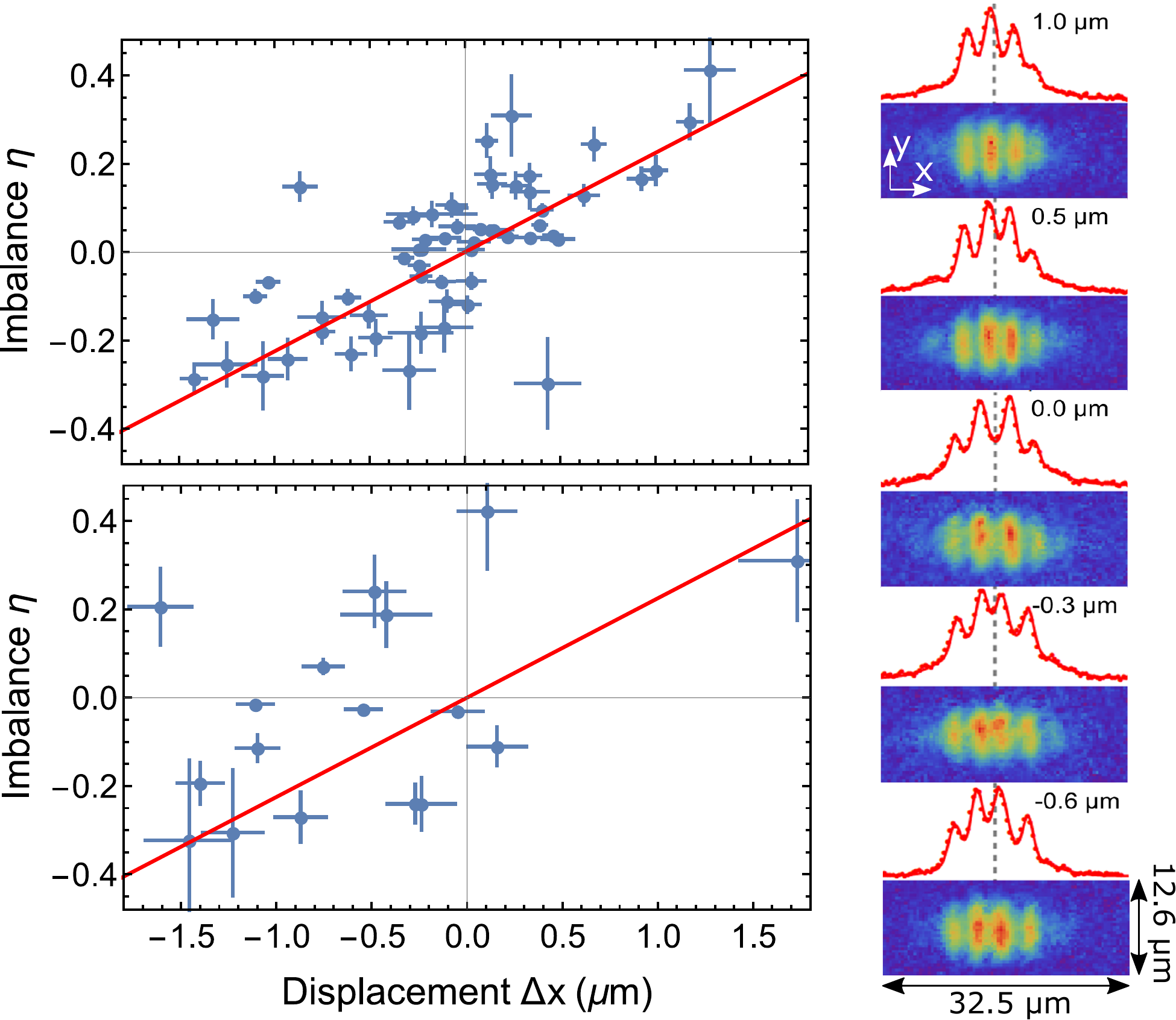}
	\put(1,82){\textbf{a}} \put(1,44){\textbf{b}} \put(73,82){\textbf{c}}
	\end{overpic}
	\caption{\textbf{Experimental correlation of the observed four-droplet states.} Similar to Fig.~\ref{figure3} in the main text, we show the experimental correlation in the supersolid ($97.6~a_0$, \textbf{a}) and isolated droplet ($91.2~a_0$, \textbf{b}) region, as well as example in-situ images of four-droplet states (\textbf{c}). For the four-droplet states we observe a clear correlations of imbalance and displacement throughout the supersolid region. 
	}
	\label{sup4_2}
\end{figure} 

In our experiment we also find a sizeable amount of realizations with four-droplets, for which we perform the same statistical analysis as we presented in the main text for the three-droplet states. The same as for the three-droplet state, we find a clear correlation for the supersolid regime (Fig.~\ref{sup4_2}\textbf{a}), while the variance of the data is larger for the isolated droplet regime (Fig.~\ref{sup4_2}\textbf{b}) indicating that no correlation exists in the isolated droplet region. This can also be seen in the example images of four-droplet states in the supersolid regime with different displacements, which are shown in Fig.~\ref{sup4_2}\textbf{c}. Similar to the three-droplet case, we can again calculate the variance of the experimental data with respect to the theoretical correlation curve as an indicator of the existence of the Goldstone mode with respect to the contact interaction strength. Similar to Fig.~\ref{figure4}\textbf{b} of the main text, the variance of the four-droplet state shown in Fig.~\ref{sup4_3} has a clear minimum in the range where we observe phase coherent droplet arrays.

\begin{figure}[t]
	\centering
	\begin{overpic}[width=0.45\textwidth]{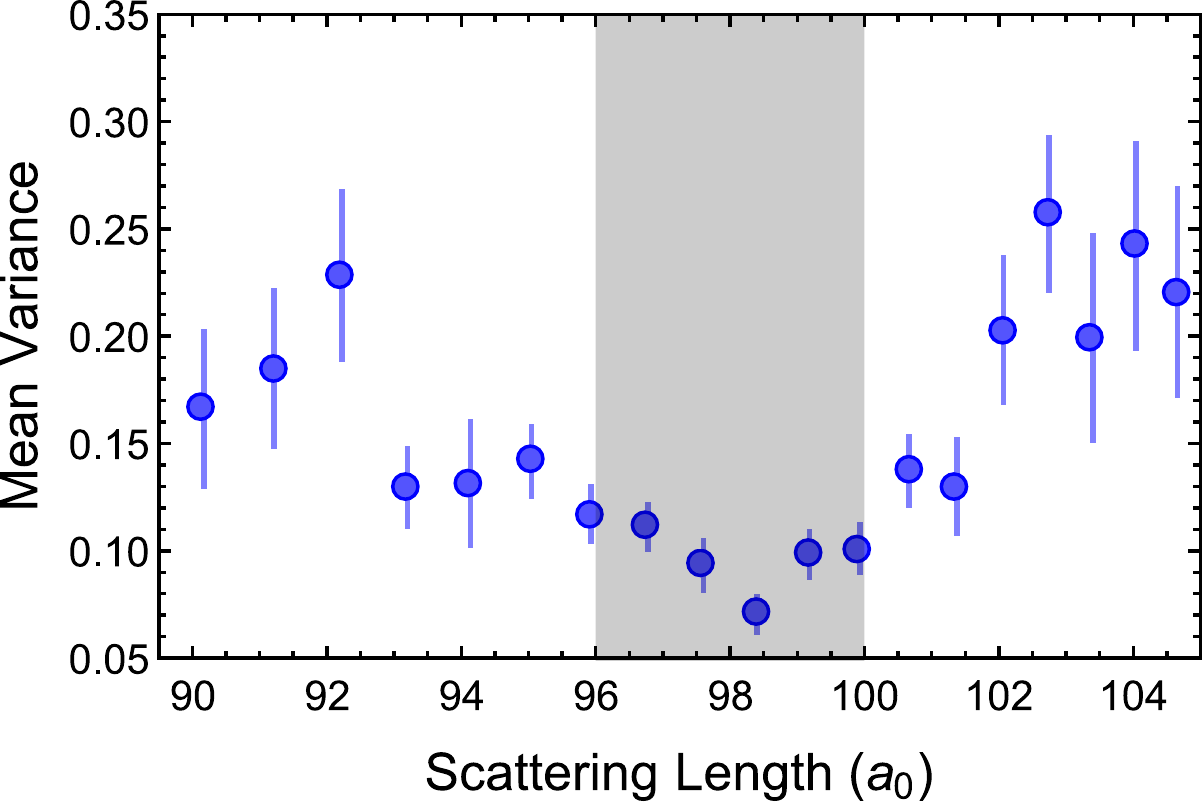}
	\end{overpic}
	\caption{\textbf{Variance of the four-droplet data with respect to the theoretical correlation.} Similar to Fig.~\ref{figure4}\textbf{b} of the main text, we find that the variance of the four-droplet data with respect to the theoretical prediction is lowest in the supersolid region. 
	}
	\label{sup4_3}
\end{figure}

\paragraph{Theory independent evaluation.} As a test that is independent of our underlying theory, we implement a linear fit to the experimental data at each scattering length and compare the data to this fit, instead of the theoretically obtained correlation. This linear fit allows us to do two complimentary tests. First, we can again calculate the spread of our experimental measurements with respect to the linear fit. In agreement with the comparison to the theoretical curve, the calculated variance with respect to the linear fit is again lowest directly in the phase-coherent regime and increases by changing the scattering length in either direction. As a second check, we can look at the absolute value of the intersection point of the linear fit with the axis of the displacement. For a supersolid state, this intersection should be at zero, corresponding to a symmetric droplet array with a vanishing imbalance at no displacement. For isolated droplets on the other hand, initial fluctuations during the formation process cannot be compensated, meaning that we can get imbalanced droplet arrays, even if the array is in the center of the cloud. The obtained intersection points for the observed three- and four-droplet states are shown in Fig.~\ref{sup5}\textbf{a} \& \textbf{b}, respectively. This shows that the intersection point is only close to zero in the supersolid region. Both of these theory independent checks act as an additional proof of the existence of a correlation between the imbalance and the displacement arising due to the low-energy Goldstone mode of the system.

\begin{figure}[t]
	\centering
	\begin{overpic}[width=0.4\textwidth]{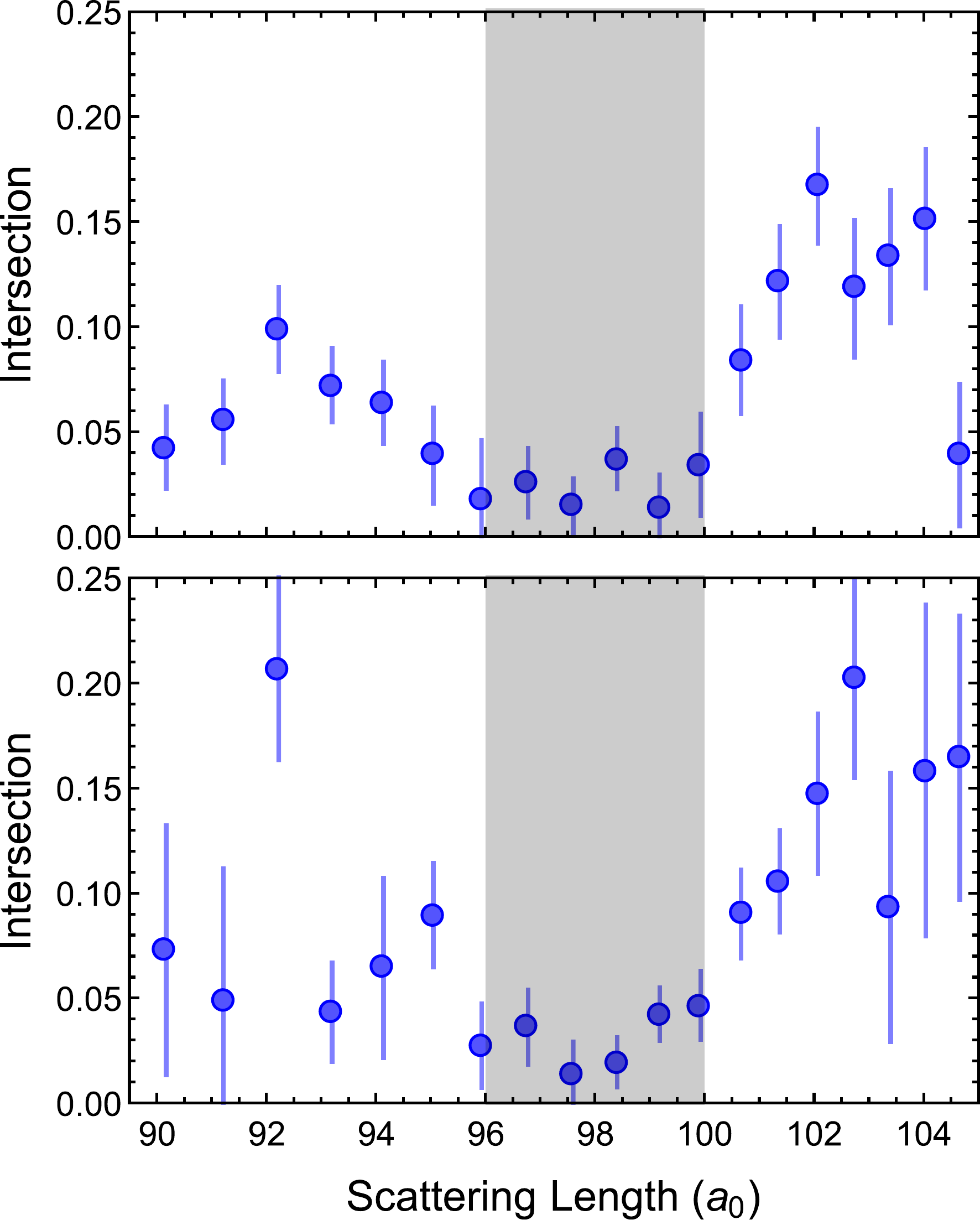}
	\put(0, 99){\textbf{a}} \put(0, 52){\textbf{b}}
	\end{overpic}
	\caption{\textbf{Theory independent evaluation.} Intersection point of the fit with the displacement axis for the three-droplet states (\textbf{a}) and the four-droplet states (\textbf{b}) obtained from a linear fit to the $\eta - \Delta x$ correlation data across the explored phase diagram. An intersection point close to zero indicates the presence of a superfluid flow that can compensate fluctuations during the formation process. The shown error bars represent the fit error of the intersection point.
	}
	\label{sup5}
\end{figure} 

\end{methods}

\bibliographystyle{naturemag}
\bibliography{reference}

\begin{thebibliography}{10}
\bibitem{footnote1}
\bibinfo{note}{Strictly speaking this is only valid for identical particles
  with no other degrees of freedom.}
\end{thebibliography}

\begin{addendum}
 \item[Acknowledgements] We acknowledge insightful discussions with the groups of F. Ferlaino, G. Modugno, L. Santos and T. Pohl, as well as with A. Pelster and A. Balaz. This work is supported by the German Research Foundation (DFG) within FOR2247 under Pf381/16-1 and Bu2247/1, Pf381/20-1, and FUGG INST41/1056-1. T.L. acknowledges support from the EU within Horizon2020 Marie Sk\l odowska Curie IF (Grants No.~746525 coolDips).
 \item[Author Contributions] M.G., F.B., J.S performed the experiment and analyzed the data. J.H. and M.W. performed the numerical analysis. H.P.B., T.L., and T.P. provided scientific guidance in experimental and theoretical questions. All authors contributed to the interpretation of the data and the writing of the manuscript.
 \item[Author Information] 
 The authors declare no competing financial interests. Correspondence and requests for materials should be addressed to T.P.~(email: t.pfau@physik.uni-stuttgart.de).
\end{addendum}

\end{document}